\documentclass[review]{elsarticle}

\usepackage{lineno,hyperref}
\usepackage{graphicx}
\usepackage{color}
\usepackage{tabularx}
\usepackage{amsmath,amssymb}

\modulolinenumbers[5]

\journal{Nuclear Physics A}

\graphicspath{{.}{figures/}}

\newcommand{\beq}{\begin{equation}}
\newcommand{\eeq}{\end{equation}}
\newcommand{\bea}{\begin{eqnarray}}
\newcommand{\eea}{\end{eqnarray}}
\newcommand{\bce}{\begin{center}}
\newcommand{\ece}{\end{center}}
\newcommand{\komma}{\;,}
\newcommand{\pkt}{\;.}
\newcommand{\half}{\frac{1}{2}}

\newcommand{\inv}[1]{\frac{1}{#1}}

\newcommand{\bac}{\begin{accentcolor}}
\newcommand{\eac}{\end{accentcolor}}

\newcommand{\ave}[1]{\left<#1\right>}

\newcommand{\Nf}{N_{\text{f}}}
\newcommand{\Nc}{N_{\text{c}}}

\newcolumntype{L}[1]{>{\raggedright\arraybackslash}p{#1}} 
\newcolumntype{C}[1]{>{\centering\arraybackslash}p{#1}} 
\newcolumntype{R}[1]{>{\raggedleft\arraybackslash}p{#1}} 

\begin{document}

\begin{frontmatter}

\title{Spectral Functions from the Functional Renormalization Group}

\author[TUD,GSI]{Jochen~Wambach\corref{mycorrespondingauthor}}
\cortext[mycorrespondingauthor]{Corresponding author.\\Talk presented at ``45 years of nuclear theory at Stony Brook - A tribute to Gerald E. Brown".}
\ead{Jochen.Wambach@physik.tu-darmstadt.de}

\author[TUD]{Ralf-Arno~Tripolt}

\author[RKU]{Nils~Strodthoff}

\author[TUD,JLU]{Lorenz~von~Smekal}

\address[TUD]{Institut f\"ur Kernphysik - Theoriezentrum,
Technische Universit\"at Darmstadt, Germany} 
\address[GSI]{GSI Helmholtzzentrum f\"ur 
Schwerionenforschung GmbH, Germany} 
\address[RKU]{Institut f\"ur Theoretische Physik,
Ruprecht-Karls-Universit\"at Heidelberg, Germany} 
\address[JLU]{Institut f\"ur Theoretische Physik,
Justus-Liebig-Universit\"at Giessen, Germany} 

\begin{abstract}
In this article we wish to present a new method  to obtain spectral 
functions at finite temperature and density from the Functional 
Renormalization Group (FRG). The FRG offers a powerful non-perturbative 
tool to deal with phase transitions in strong-interaction matter under 
extreme conditions and their fluctuation properties. Based on a 
thermodynamically consistent truncation we derive flow equations 
for pertinent two-point functions in Minkowski space-time. 
We demonstrate the feasibility of the method by calculating mesonic 
spectral functions in hot and dense hadronic matter using the 
quark-meson model as a simple example.

\end{abstract}


\end{frontmatter}


\section{Introduction}

The in-medium modifications of hadron properties have been dear to 
Gerry's  heart and his work on this subject has made him and his 
collaborators major drivers of this field for several decades \cite{Brown1991,Brown2004}.
The discussions are far from over and we wish to pay tribute to Gerry's
physical insights by adding a new and promising alternative for computing 
real-time spectral functions in hot and dense QCD matter. The aim is to 
set up a framework that goes beyond mean-field theory (which Gerry employed 
very successfully during his career) by incorporating quantum fluctuations. 
This is particularly important for the understanding of the restoration of 
broken chiral symmetry in the hadronic medium. 

Spectral functions encode information on the particle spectrum as well as 
collective excitations of a given system. A thermodynamically consistent calculation 
beyond the Hartree-Fock level of such real-time observables represents an 
inherently difficult problem. Although mean-field calculations might capture 
the gross features of the equilibrium properties, quantitative predictions 
and correct descriptions of critical phenomena require the proper inclusion 
of fluctuations.  Several self-consistent methods are available among which 
the Functional Renormalization Group (FRG) is particularly suitable and widely 
used in quantum field theory and condensed matter physics 
\cite{Berges:2000ew,Polonyi:2001se,Pawlowski:2005xe,Schaefer:2006sr,Kopietz2010,Braun:2011pp,Gies2012}.

A technical difficulty which is common to all Euclidean approaches to Quantum 
Field Theory is the need to analytically continue from imaginary to real time 
especially for dynamic processes with time-like momentum transfers. At finite 
temperature these continuations are often based on numerical (i.e. noisy) data 
at discrete Matsubara frequencies and several approximate methods for the 
reconstruction of real-time spectral functions have been used 
\cite{Vidberg:1977,Jarrell:1996,Asakawa:2000tr,Dudal2013} with varied success.
Therefore any approach that can deal with the analytic continuation explicitly 
is highly desirable. Such alternative approaches have been proposed in 
\cite{Strodthoff:2011tz, Kamikado2013} and \cite{Floerchinger2012}. They involve 
an analytic continuation on the level of the FRG flow equations for two-point 
correlation functions and have been applied in \cite{Kamikado2013} 
to a model system with O(4) internal symmetry in vacuum. 
More recently, this method has been applied to obtain spectral functions from
the quark-meson model at finite temperature and density \cite{Tripolt2014}.
These studies form the basis for the present discussion of hadronic matter at 
high temperatures and large baryo-chemical potentials.

\section{The Functional Renormalization Group}

To set the stage we start out with the basic ideas of the FRG, by discussing 
the effective potential of a system in thermal equilibrium, the ensuing 
correlation functions and their flow equations.

\subsection{Effective action}

The principal object of statistical physics is the (Euclidean) partition 
function $Z$ from which the equilibrium properties of a thermal system can 
be derived. $Z$ can be represented as a Feynman path integral (for simplicity 
in a real field variable $\phi$). In the presence of an external source $j$ it reads:
\beq
Z[j]=e^{W[j]}=\int\!\!\left[{\cal D}\phi\right]\;e^{-S[\phi]+\int\! d^4x\; \phi(x)j(x)}\komma
\eeq
and  $W[j]$ generates all $n$-point Green functions via functional derivatives 
w.r.t. the source function $j$. In particular  
\beq
\left. \frac{\delta W[j]}{\delta j(x)}\right|_{j=0}=\ave{\phi(x)}\equiv \varphi(x)
\eeq 
and 
\beq
\left. \frac{\delta^2 W[j]}{\delta j(x)\delta j(y)}\right|_{ j=0}=
\ave{\phi(x)\phi(y)}-\ave{\phi(x)}\ave{\phi(y)}\equiv
G(x,y)\pkt
\eeq
The central object of the FRG is the effective action $\Gamma[\varphi]$ as 
the Legendre transform of $W[j]$:
\beq
\Gamma[\varphi]=-W[j]+\int\! \!d^4x\; \varphi(x)j(x)\pkt
\eeq
It is stationary at the field minimum $\varphi_0$ and related to the (grand) 
canonical potential:
\beq
\left. \frac{\delta\Gamma[\varphi]}{\delta\varphi}\right|_{\varphi=\varphi_0}
=0;\quad\to\quad \Omega(T,\mu)=\frac{T}{V} \Gamma[\varphi_0]\komma
\eeq
where $T$ denotes the temperature and $\mu$ the chemical potential of the system.

\subsection{Wilsonian coarse graining}

Wilson's coarse graining \cite{Wilson1971, Wilson1974} starts with a choice of resolution scale $k$ and 
formally splits the field variable $\phi(x)$ into low- and high-frequency modes:
\beq
\phi(x)=\phi_{q\leq k}(x)+\phi_{q>k}(x)\pkt
\eeq
This allows to rewrite the partition function $Z$ in the following way:
\beq
Z[j]=\int\![{\cal D}\phi]_{q\leq k} \underbrace{\int\![{\cal D}\phi]_{q>k}\;
e^{-S[\phi]+\int\! d^4x\; \phi(x) j(x)}}_{=Z_k[j]}\komma
\eeq
thus introducing a scale-dependent partition function $Z_k[j]$ which becomes 
the full partition function in the infra-red limit:
\beq
\lim_{k\to 0}Z_k[j]=Z[j]\pkt
\eeq
To render $Z$ ultra-violet and infra-red finite a regulator function $R_k(q)$ 
is introduced (whose specific form is irrelevant) with the boundary conditions
\beq
\lim_{k\to 0}R_k(q)=0\quad \lim_{k\to \Lambda}R_k(q)=\infty\komma
\eeq
where $\Lambda$ denotes an ultra-violet cut-off scale. The regulator function 
adds an additional term $\Delta S_k[\phi]$ to the action such that 
\beq
Z_k[j]=\int\![{\cal D}\phi]\; e^{-S[\phi]-\Delta S_k[\phi]+\int\! d^4x\; \phi(x)j(x)}
\eeq
with
\beq
\Delta S_k[\phi]=\half \int\! \!\frac{d^4q}{(2\pi)^4}\;\phi(-q)R_k(q)\phi(q)
\eeq
formally acting as a $k$-dependent mass term. For the scale-dependent effective 
action $\Gamma_k[\varphi]$ this implies
\beq
\Gamma_k[\varphi]=-\ln Z_k[j]+\int\!\! d^4x\; \varphi(x) j(x)-\Delta S_k[\varphi]\komma
\eeq
which interpolates between $k=\Lambda$ where no fluctuations are considered 
(classical action) and the full quantum action at $k=0$:  
\beq
\lim_{k\to \Lambda}\Gamma_k[\varphi]=S[\varphi];\quad\lim_{k\to 0}
\Gamma_k[\varphi]=\Gamma[\varphi]\pkt
\eeq

\subsection{Flow equations}
\begin{figure}[h]
\centering\includegraphics[width=0.45\columnwidth]{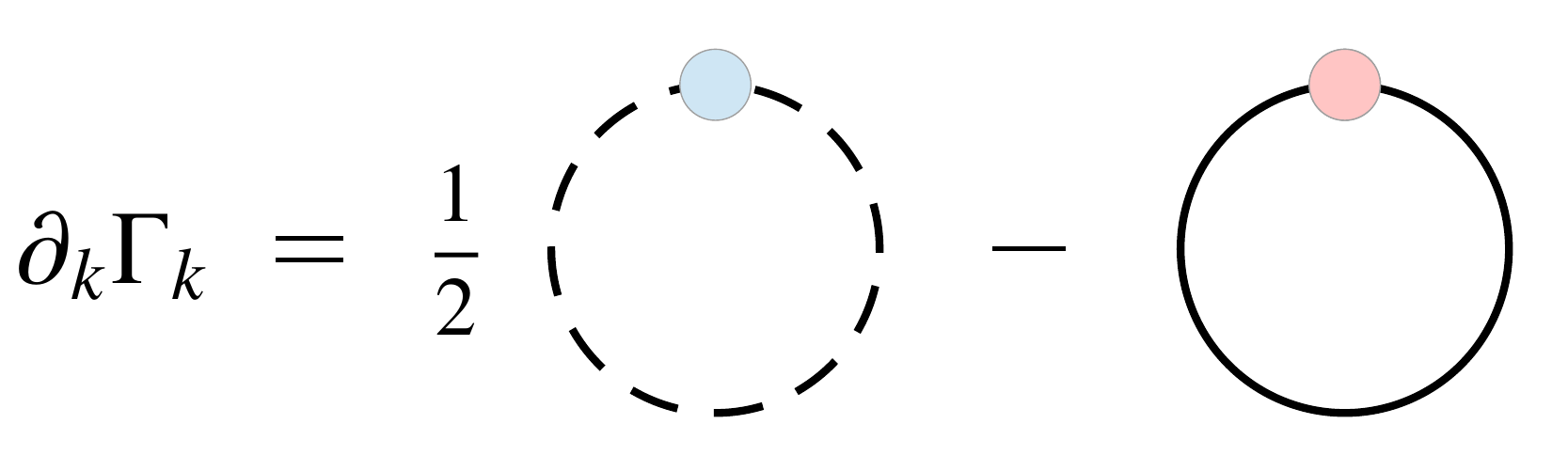}
\caption{(color online) Diagrammatic representation of 
the flow equation for the effective action. Dashed (solid) lines 
represent bosonic (fermionic) propagators and circles represent regulator 
insertions $\partial_k R_k$.}
\label{fig:flow_Gamma} 
\end{figure}

The scale-dependence of the effective action $\Gamma_k$ is governed by 
an exact flow equation \cite{Wetterich:1992yh},
\beq
\partial _{k}\Gamma _{k}[\varphi]=\frac{1}{2}\,{\rm Tr}\left(\partial_{k}R_{k} 
\left[ \Gamma_{k}^{(2)}+R_{k}\right]^{-1}\right)\komma
\label{eq:wetterich} 
\eeq 
where the trace involves integration of internal momenta as well as summations over 
internal indices. The flow equation involves the propagator $\Gamma^{(2)}_k$ 
as a second functional derivative,
\beq
\Gamma_{k}^{(2)}(q)=\frac{\delta^2\Gamma_k[\varphi]}{\delta\varphi(-q)\delta\varphi(q)}\komma
\eeq
which is graphically represented in Fig.~\ref{fig:flow_Gamma} by a one-loop 
equation for theories involving both bosonic (dashed line) and fermionic 
(full line) degrees of freedom.

\section{The O(4) model}
For illustrative purposes and to introduce the approximations involved in the 
analytic continuation for real-time spectral functions we start with a simple 
model system, the O(4) linear-sigma model, involving an iso-triplet of pions 
and the sigma meson:
\beq
\varphi=(\varphi_1,\dots,\varphi_4)=(\sigma,\vec\pi)\pkt
\eeq
For the vacuum, results for the sigma and the pion spectral functions have 
been presented in \cite{Kamikado2014}. The starting point is the effective 
action of the O(4) model. Even though the exact flow equations can be written 
down concisely, they are functional equations that are difficult to solve. 
For practical applications one has to resort to truncations. In a simple 
possibility, which has proven quite successful in many instances, one is 
guided by an expansion of $\Gamma_k$ in the number of space-time derivatives. 
To lowest order only the potential $U$ of a given Lagrangian becomes 
scale dependent (local potential approximation (LPA)) and the corresponding O(4)-action reads:  
\beq
\Gamma_k[\varphi]=\int\!\! d^4x\left\{\half(\partial_\mu\varphi)^2+U_k(\varphi^2)
-c\sigma\right\};\quad \varphi^2=\varphi_i\varphi^i=\sigma^2+\vec\pi^2\pkt
\label{eq:ONGamma}
\eeq
Using the Wetterich equation, Eq. (\ref{eq:wetterich}), one easily obtains the 
following flow equation for the effective potential,
\beq
\partial_k U_k=I_\sigma+3I_\pi;\quad I_i=\half{\rm Tr_q}\left(\partial_{k}R_{k}(q) 
\left[ \Gamma_{k,i}^{(2)}(q)+R_{k}(q)\right]^{-1}\right)\pkt
\label{eq:Is}
\eeq
For a simple choice of the regulator function $R_k(q)$,
\beq
R_k(q)=(k^2-\vec{q}^{\,2})\Theta(k^2-\vec{q}^{\,2})\komma
\label{eq:reg}
\eeq
the loop functions right-hand side of the flow equation takes a simple analytic form,
\beq
I_i=\frac{k^4}{3\pi^2}\inv{2E_i}\komma
\eeq
with 
\beq
E_\pi=\sqrt{k^2+2U'}; \quad E_\sigma=\sqrt{k^2+2U'+4U''\varphi^2};\quad U'=
\frac{\partial U}{\partial \varphi^2}\; \textrm{etc.}
\eeq
In order to solve the flow equation for the effective potential, Eq. (\ref{eq:Is}), 
the radial field component $\sigma$ is discretized along a one-dimensional grid
in field space while the angular field components are 
set to their expectation value, $\langle \vec{\pi}\rangle=0$.
In this way the original partial differential equation turns into a set of ordinary differential equations
which can be solved numerically \cite{Schaefer:2004en}.

\begin{figure}[t]
\centering\includegraphics[width=0.99\columnwidth]{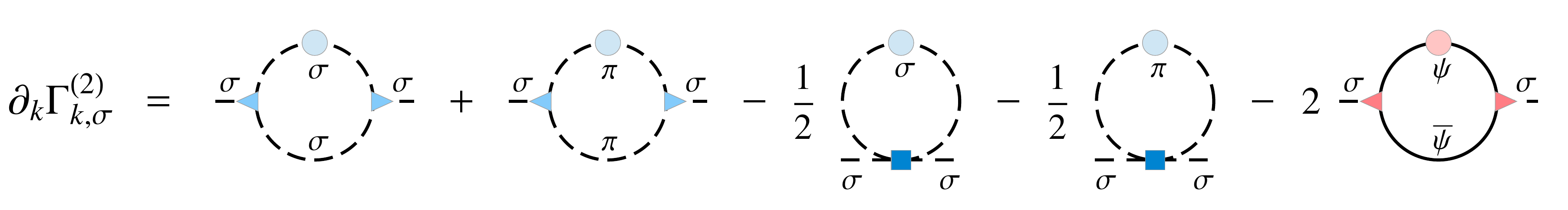}\\
\centering\includegraphics[width=0.99\columnwidth]{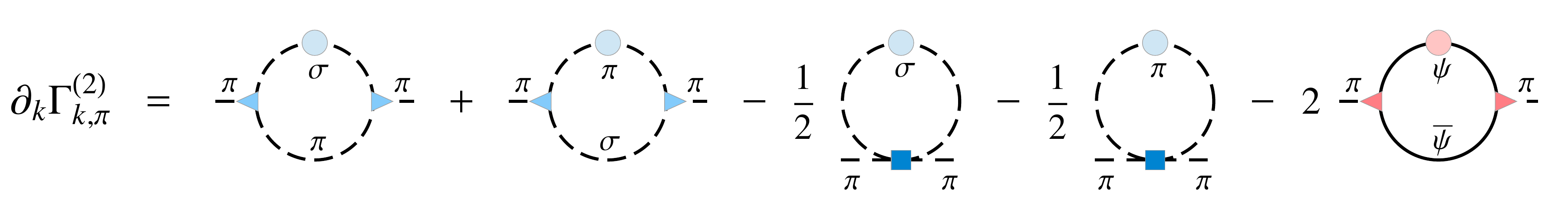}
\caption{(color online) Diagrammatic representation of 
the flow equation for the sigma and pion 2-point functions for the quark-meson model. 
In the case of the O(4) model the quark loop does not appear.}
\label{fig:flow_Gamma2} 
\end{figure}

Taking two functional derivatives of the flow equation for $\Gamma_k$, Eq. (\ref{eq:ONGamma}), 
yields the flow equations for the inverse propagators, which are graphically depicted in 
Fig. \ref{fig:flow_Gamma2}.\footnote{Explicit expressions for the flow equations are given in \cite{Kamikado2014}.}
These require knowledge of 3- and 4-point vertex functions which we obtain from the
corresponding momentum independent but scale dependent couplings of the effective
average action in the LPA.
We emphasize that our truncation is thermodynamically consistent
in the sense that the static screening masses obtained from the 2-point functions 
agree with the masses obtained from the curvature of the effective potential, 
as can be seen from the flow equations which satisfy
\bea
\partial_k\Gamma^{(2)}_{k,\pi}(p=0)&=&2\partial_k U'_k\komma\\
\partial_k\Gamma^{(2)}_{k,\sigma}(p=0)&=&2\partial_k U'_k+4\partial_k U''\varphi^2\pkt
\eea
This also obviates the symmetry-preserving character of the truncation in that it yields the pion as a 
Nambu-Goldstone boson in the chiral limit ($c=0$).

In order to obtain the sigma and pion spectral functions, we calculate the 
retarded inverse propagator by analytically continuing the flow equations 
from imaginary to real time,
\bea
\Gamma^{(2) R}_{k,j} (\omega) = -\lim_{\epsilon\rightarrow 0} \Gamma^{(2) E}_{k,j}
(p_0=i\omega- \epsilon))
\, , \quad j=\pi,\sigma \komma
\label{eq:analytic_continuation}
\eea
where the real parameter $\epsilon$ is kept small but finite in the numerical calculations. 
The resulting flow equations for the retarded 2-point functions are then solved 
using the scale-dependent (but momentum independent) 3- and 4-point vertices as 
well as quasi-particle masses and energies extracted from the LPA result for 
the scale-dependent effective average action.
Moreover, the radial field component $\sigma$ is set to its expectation value in 
the infra-red, as determined from the global minimum of the effective potential.
Finally, the spectral functions are given by
\beq
\label{eq:spectralfn} \rho_j(\omega) = \frac{1}{\pi} \frac{{\rm
Im } \, \Gamma^{(2) R}_{j}(\omega)}{\left({\rm Re}\, \Gamma^{(2)R}_{j}
(\omega)\right)^2 +\left({\rm Im}\, \Gamma^{(2) R}_{j}(\omega)\right)^2}
\, , \quad j=\pi,\sigma\pkt
\eeq

\subsection{Spectral functions in the vacuum}

The following results for the sigma and pion spectral functions were obtained 
using the parameter sets given in Table \ref{tab:parameter}, where the potential 
was chosen as $U_{k=\Lambda}=a\phi^2+b\phi^4$ in the UV.

\begin{table}[t]
\centering
\begin{tabular}{C{1.3cm}|C{1.3cm}|C{1.1cm}||C{1.3cm}|C{1.3cm}|C{1.3cm}}
 $a/\Lambda^2$ & $b$ & $c/\Lambda^3$ & $f_\pi$ &$m^\mathrm{scr}_{\pi}$&$m^\mathrm{scr}_{\sigma}$ \\
\hline
-0.30&3.65&0.014&93.0&137.2&425.0\\
\hline
-0.34&3.40&0.002&93.1&16.4&299.8\\
\end{tabular}
\caption{Parameter sets for a UV cutoff $\Lambda=500$~MeV corresponding to two 
different pion masses. The physical parameters, $f_{\pi}$ and the meson masses, 
are given in MeV.}
\label{tab:parameter}
\end{table}

\begin{figure}[h]
\begin{center}
\includegraphics[width=0.47\columnwidth]{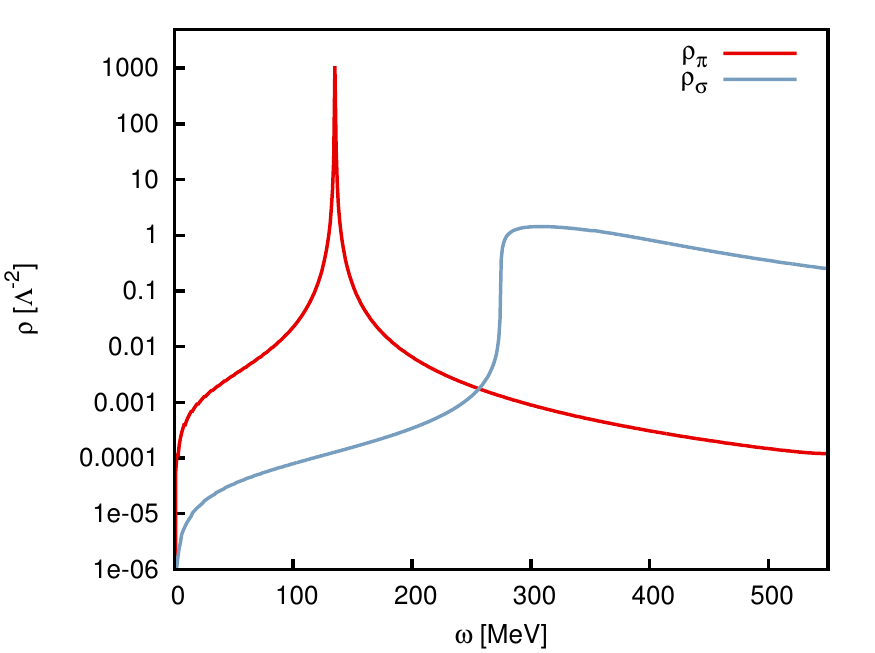}\hspace{0.04\columnwidth}
\includegraphics[width=0.47\columnwidth]{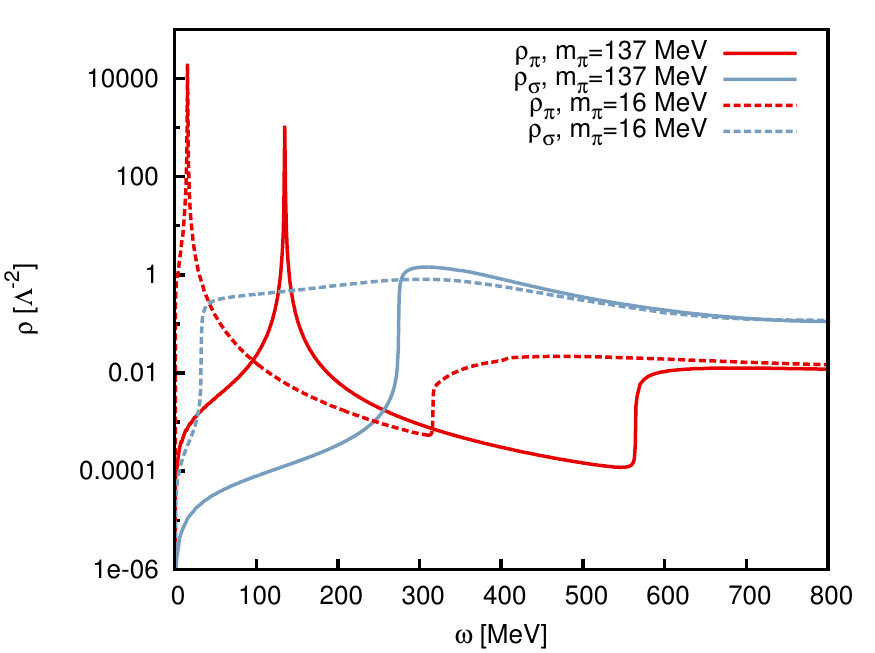}
\caption{(color online) Pion ($\rho_\pi$) and sigma meson ($\rho_\sigma$) spectral functions from \cite{Kamikado2014} for two different parameter sets.}
\label{fig:spectral_ON} 
\end{center}
\end{figure}

In Fig. \ref{fig:spectral_ON} we show the spectral functions for physical pion 
mass, $m_\pi=137$ MeV, (left) as well as near the chiral limit $m_\pi=16$ MeV 
(right). For the physical mass, the pion spectral function exhibits a sharp peak 
at 135~MeV, as one would expect, while the sigma spectral function starts at the 
2-pion threshold with a sharp increase in the spectral density, followed by a 
broad maximum at about 312~MeV above this threshold. Although, in the 
ultra-violet, the sigma meson is sharp it acquires a large decay width during the $k\to 0$ evolution from the decay into two pions.

When approaching the chiral limit, one expects the spectral weight of the pion pole to increase more and more as this pole moves closer to the $\omega=0$ axis, where it eventually accumulates the full spectral weight. The spectral sum rule implies that all other contributions to the spectral function should decrease with decreasing pion mass. Both trends are seen in Fig.~\ref{fig:spectral_ON} where
we have extended the frequency range to include the $\pi^* \rightarrow \sigma \pi$ threshold in the pion spectral function.

\section{In-medium spectral functions}

We now apply the method, presented above, to strong-interaction matter within the quark-meson model for finite temperature and density \cite{Tripolt2014}. The quark-meson model serves as a low-energy effective model for QCD with $\Nf=2$ light quark flavors, which shares chiral symmetry and its breaking pattern. The scale-dependent effective average action (in the LPA) reads as follows:
\beq
\Gamma_{k}[\bar{\psi} ,\psi,\varphi]= \int\! d^{4}x \Big\{
\bar{\psi} \left({\partial}\!\!\!\slash +
h(\sigma+i\vec{\tau}\cdot\vec{\pi}\gamma_{5}) -\mu \gamma_0 \right)\psi
+\frac{1}{2} (\partial_{\mu}\varphi)^{2}+U_{k}(\varphi^2) - c \sigma 
\Big\}\pkt
\label{eq:QM}
\eeq
Using the Wetterich equation, Eq. (\ref{eq:wetterich}), with optimized regulator functions 
for bosonic and fermionic fields, cf. Eq. (\ref{eq:reg}),
the flow equation for the effective potential becomes
\beq
\partial_k U_k=I_\sigma+3I_\pi-\Nc\Nf I_\psi,
\eeq
with the loop functions $I_i$ as defined in Eq. (\ref{eq:Is}). As in the case of the O(N) model,
the flow equation for the effective potential is solved using the grid method.

The flow equations for the two-point functions are obtained by taking two functional 
derivatives of Eq. (\ref{eq:QM}) and are represented diagrammatically in Fig.~\ref{fig:flow_Gamma2}.
Therein, the quark-meson 3-point vertices are taken to be momentum and scale-independent,
$\Gamma^{(2,1)}_{\bar\psi\psi\sigma}=h$ and $\Gamma^{(2,1)}_{\bar\psi\psi\vec\pi}=ih\gamma^5\vec\tau$, while
the mesonic vertices are scale-dependent and given by appropriate derivatives of the effective potential, 
as in the case of the O(4) model.

In order to obtain the flow equations for the retarded two-point functions, we have to perform an 
analytic continuation from imaginary to real frequencies. Since we are now working at finite temperature,
this analytic continuation is not as straightforward as in the vacuum.
However, it can be shown that the following two-step procedure obeys the correct (Baym-Mermin) 
boundary conditions and yields the proper retarded propagators \cite{Baym1961,Landsman1987}.
Once the sum over the Matsubara frequencies in the flow equations are performed, we first treat the external energy as an imaginary and discrete quantity, $p_0=i\,2\pi n T$, and exploit the periodicity of the bosonic and fermionic occupation numbers which appear in the flow equations:
\beq
n_{B,F}(E+i p_0)\rightarrow n_{B,F}(E)\pkt
\eeq
For explicit expressions of the flow equations we refer to \cite{Tripolt2014}.
In a second step, we replace the discrete imaginary external energy by a continuous real energy in order to
obtain the retarded 2-point functions, cf. Eq.~(\ref{eq:analytic_continuation}).

As described for the O(4) model above, the flow equations for the real and imaginary parts of the retarded 2-point functions 
are then solved at the global minimum of the effective potential in the IR. Further details on our numerical
implementation are given in \cite{Tripolt2014}.

\subsection{Results}

\begin{table}[t]
\centering
\begin{tabular}{C{1.3cm}|C{1.1cm}|C{1.1cm}|C{1.1cm}||C{1.1cm}|C{1.1cm}|C{1.1cm}|C{1.1cm}}
$m_\Lambda/\Lambda$ & $\lambda_\Lambda$ & $c/\Lambda^3$ &  $h$& $f_\pi$ 
&$m^\mathrm{scr}_{\pi}$&$m^\mathrm{scr}_{\sigma}$&$m^\mathrm{scr}_{q}$ \\
\hline
0.794 & 2.00 & 0.00175 & 3.2 & 93.5 & 138 & 509 & 299 \\
\end{tabular}
\caption{Parameter set for the quark-meson model with $\Lambda=1000$~MeV. The pion decay constant 
as well as the particle masses are given in MeV.}
\label{tab:parameters} 
\end{table}

\begin{figure}[t]
\centering\includegraphics[width=0.5\columnwidth]{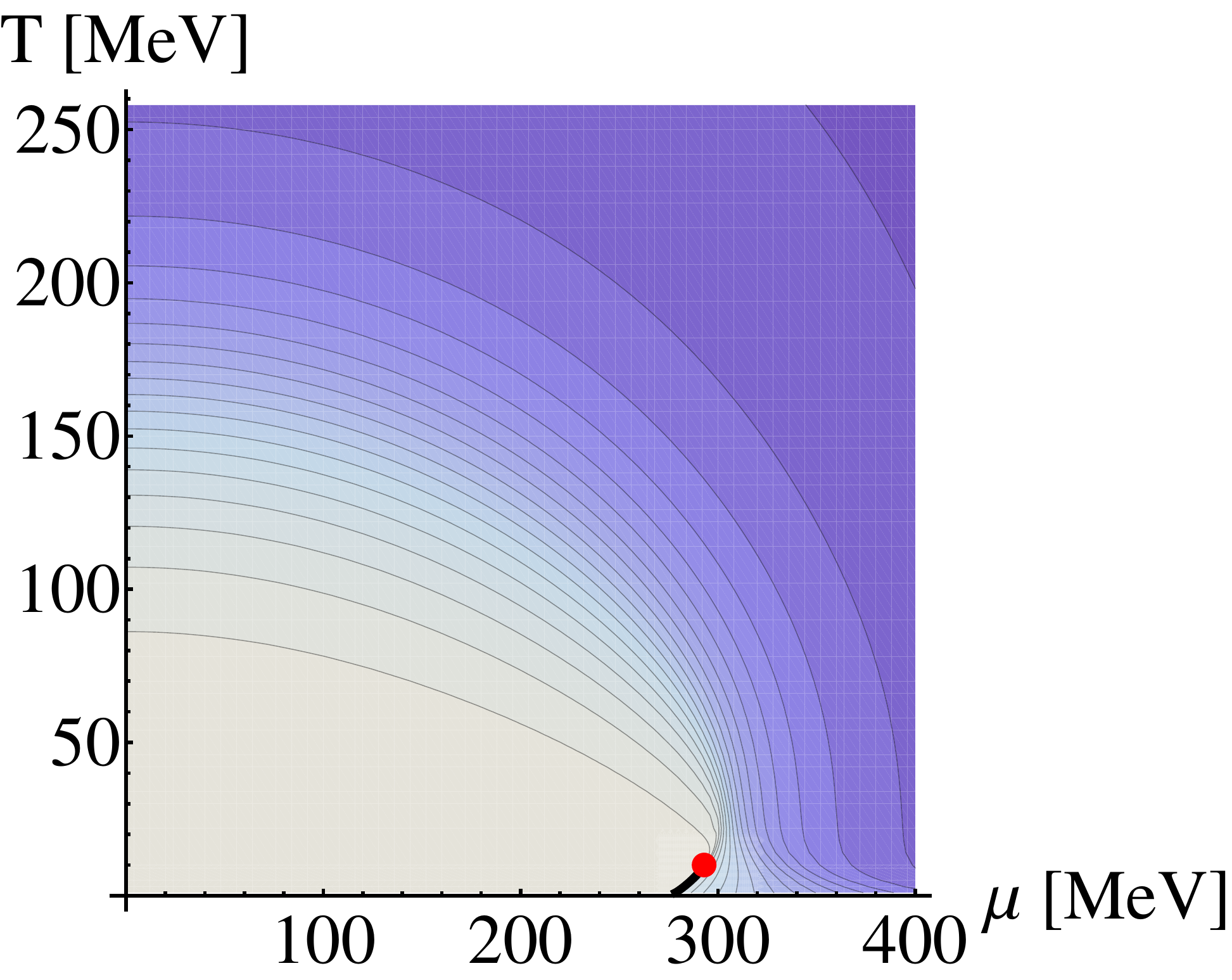}\llap{\makebox[2cm][l]{\raisebox{2cm}{\includegraphics[height=2.5cm]{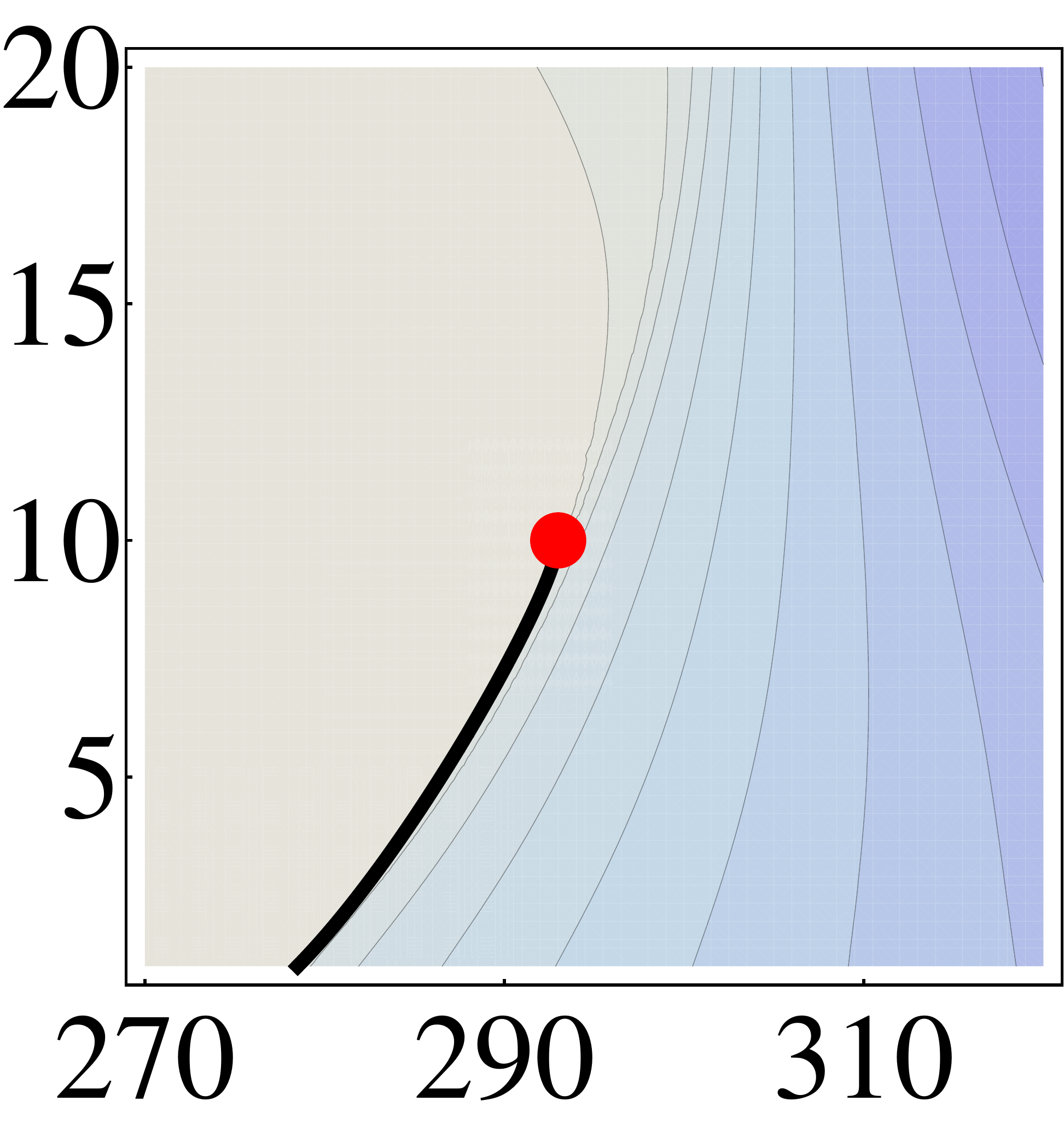}}}}
\caption{(color online) The phase diagram of the quark-meson model from 
\cite{Tripolt2014}, as obtained by using the 
parameter set in Tab. \ref{tab:parameters}, is illustrated by a contour 
plot of the magnitude of the chiral order parameter, $\sigma_0\equiv f_\pi$, 
vs. quark chemical potential~$\mu$ and temperature~$T$. The order parameter 
decreases towards higher $\mu$ and $T$, as indicated by a darker color.}
\label{fig:phase_diagram} 
\end{figure}

\begin{figure*}[t]
\includegraphics[width=0.47\columnwidth]{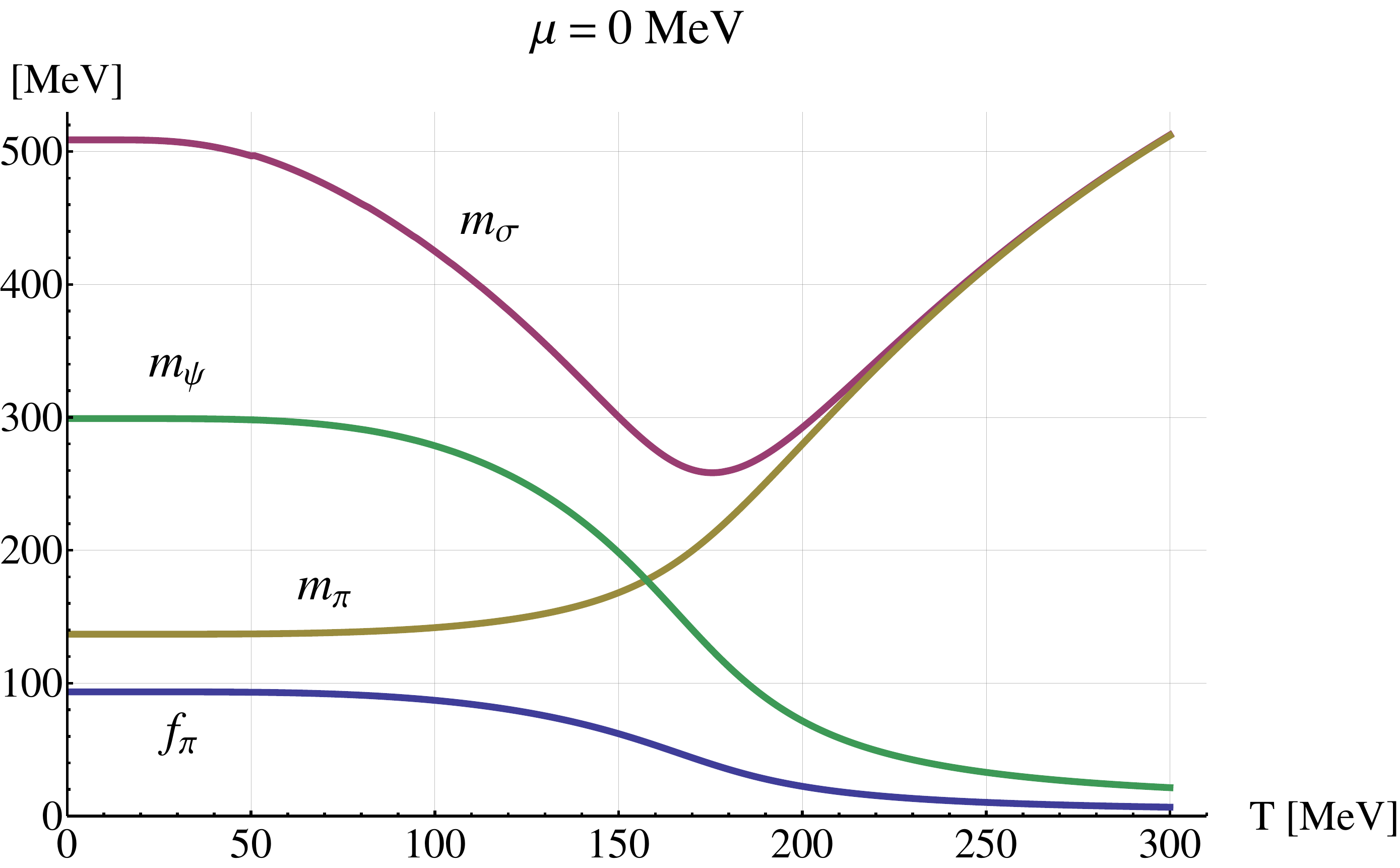}\hspace{0.04\columnwidth}
\includegraphics[width=0.47\columnwidth]{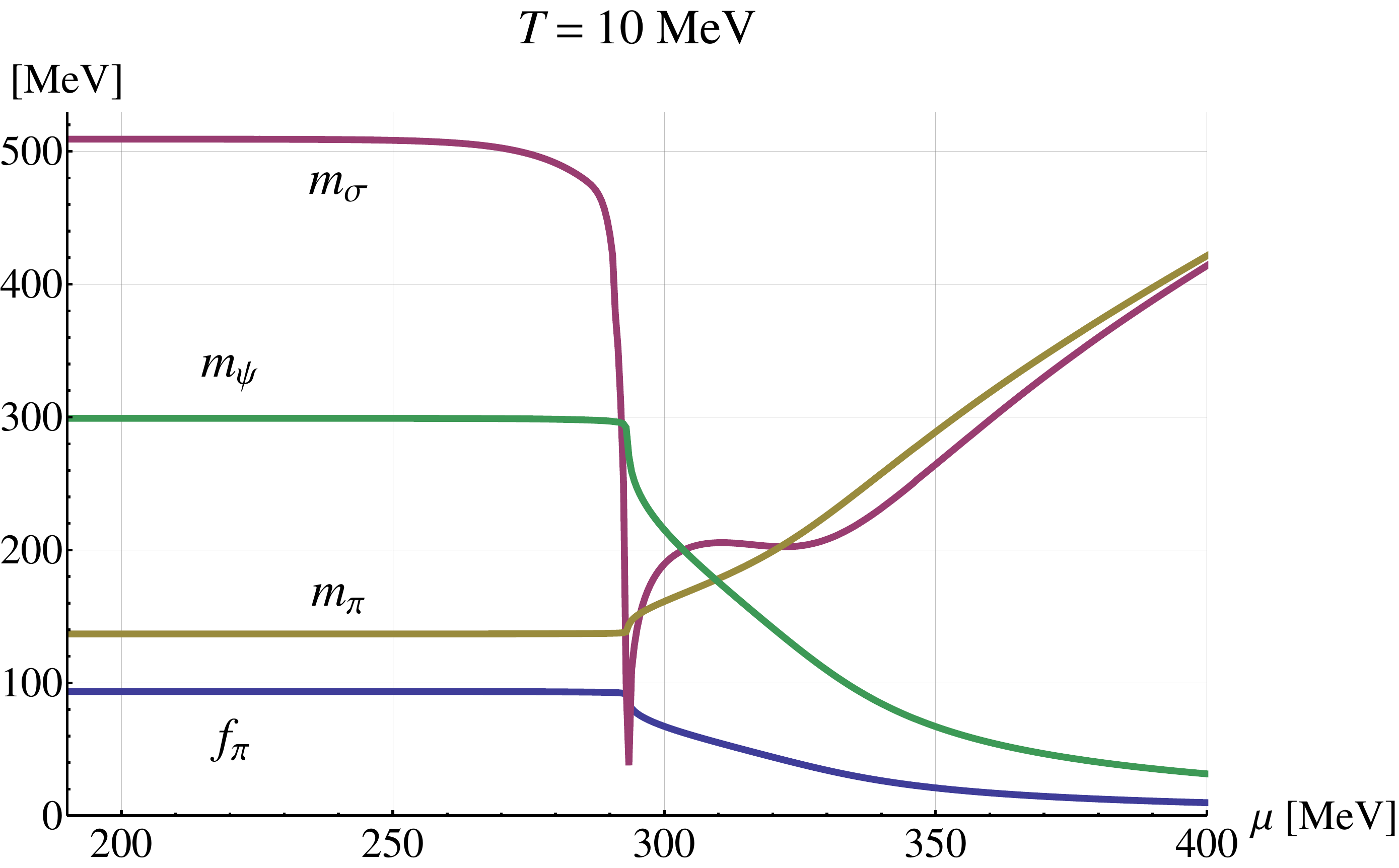}
\caption{(color online) The meson screening masses, the quark mass and the chiral order parameter, 
$\sigma_0\equiv f_\pi$, are shown vs. temperature~$T$ at $\mu=0\,{\rm MeV}$ (left panel), and 
vs. quark chemical potential~$\mu$ at $T=10\,{\rm MeV}$ (right panel), taken from \cite{Tripolt2014}.}
\label{fig:masses} 
\end{figure*}

The results for the quark-meson model were obtained using the parameter set given in 
Tab. \ref{tab:parameters}, with the UV potential chosen as
\beq
\label{eq:pot_UV} 
U_\Lambda(\phi^{2}) =
\tfrac{1}{2}m_\Lambda^{2}\phi^{2} +
\tfrac{1}{4}\lambda_\Lambda(\phi^{2})^{2}\pkt
\eeq
We first briefly discuss the resulting phase diagram as it serves as input for the calculation of 
the spectral functions. It is shown in Fig.~\ref{fig:phase_diagram} and 
displays the typical shape \cite{Schaefer:2004en} found in quark-meson 
model calculations beyond the mean-field approximation with a critical 
endpoint at $\mu=293\,{\rm MeV}$ and $T=10\,{\rm MeV}$ for this parameter set.\footnote{The low value for the chiral endpoint 
should not be of concern, since its location strongly depends on parameter choices and effects from Polyakov-loop extensions of the model.}
In addition, it is also instructive to consider the temperature- and chemical potential-dependence of the meson screening masses as they will be used to identify thresholds in the spectral functions,
cf. Fig.~\ref{fig:masses}.

\begin{figure*}
\includegraphics[width=0.45\columnwidth]{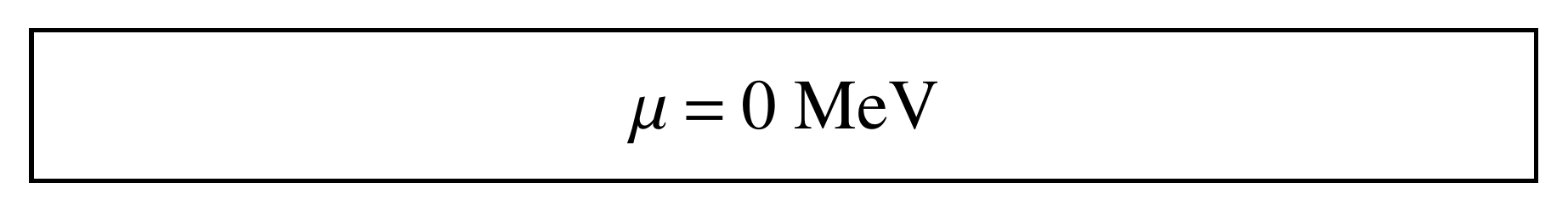}\hspace*{9mm}\includegraphics[width=0.45\columnwidth]{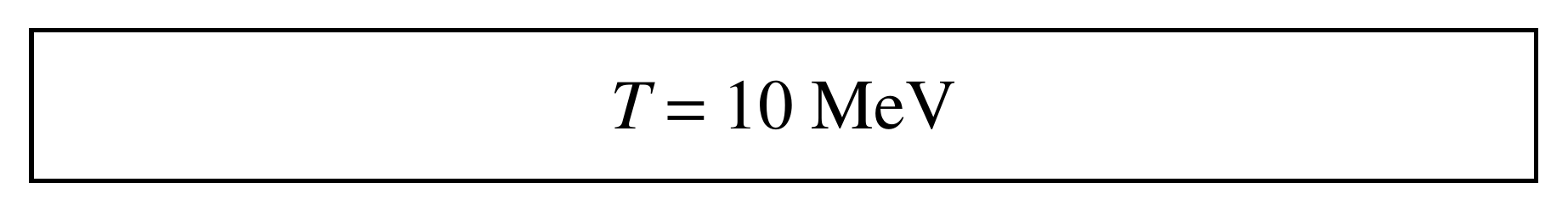}\vspace{1mm}
\includegraphics[width=0.48\columnwidth]{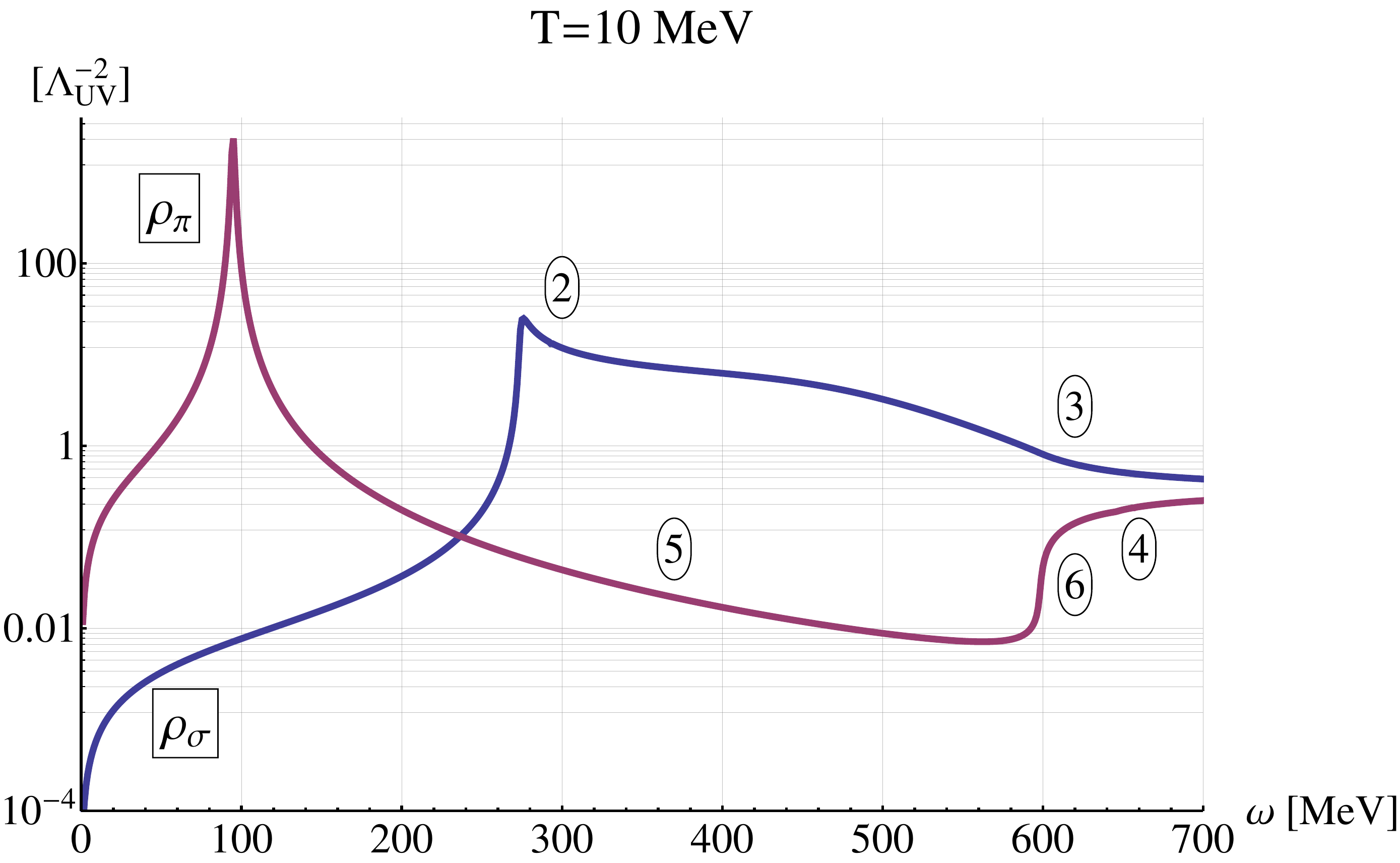}\hspace{5mm}
\includegraphics[width=0.48\columnwidth]{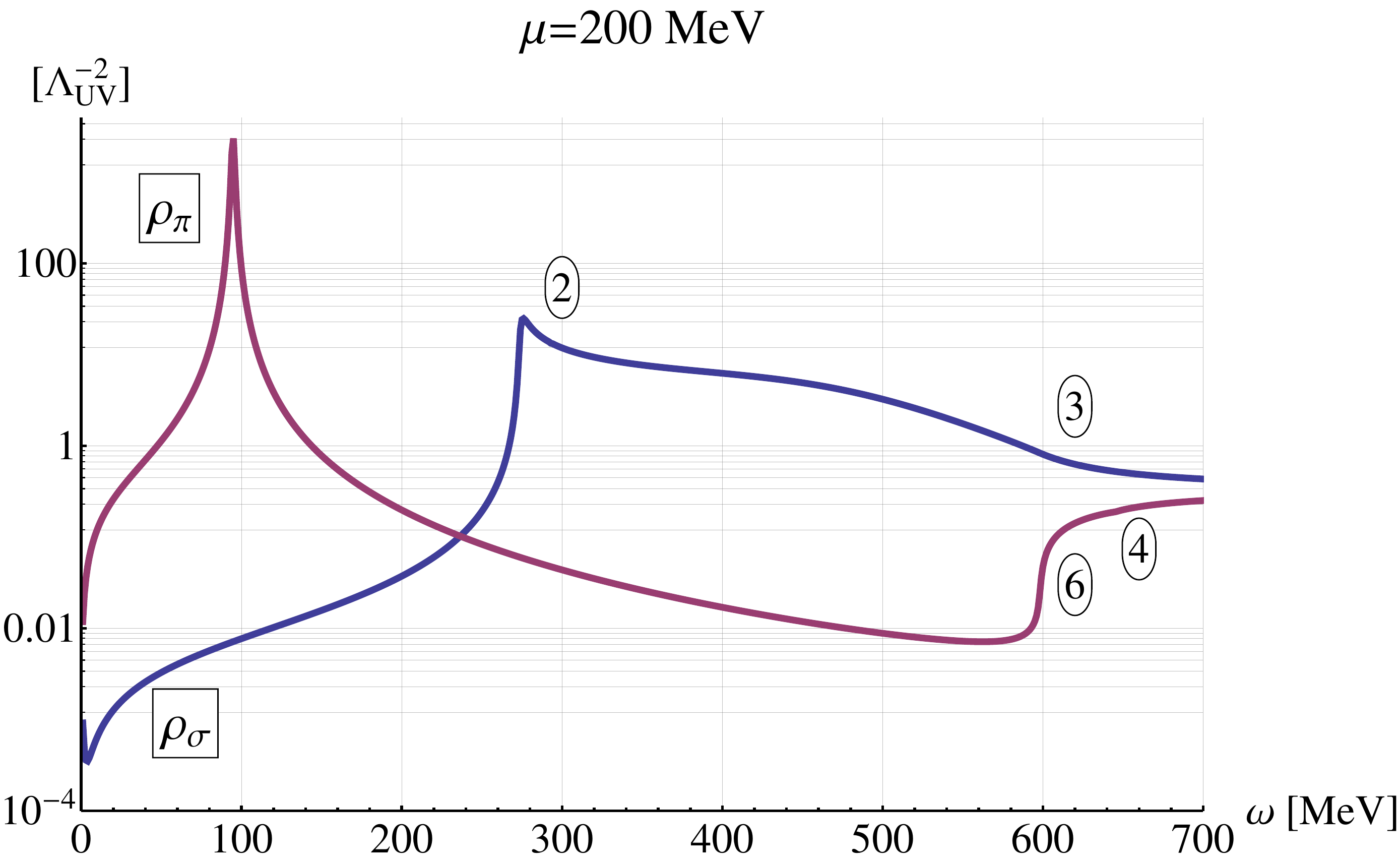}\vspace{3mm}
\includegraphics[width=0.48\columnwidth]{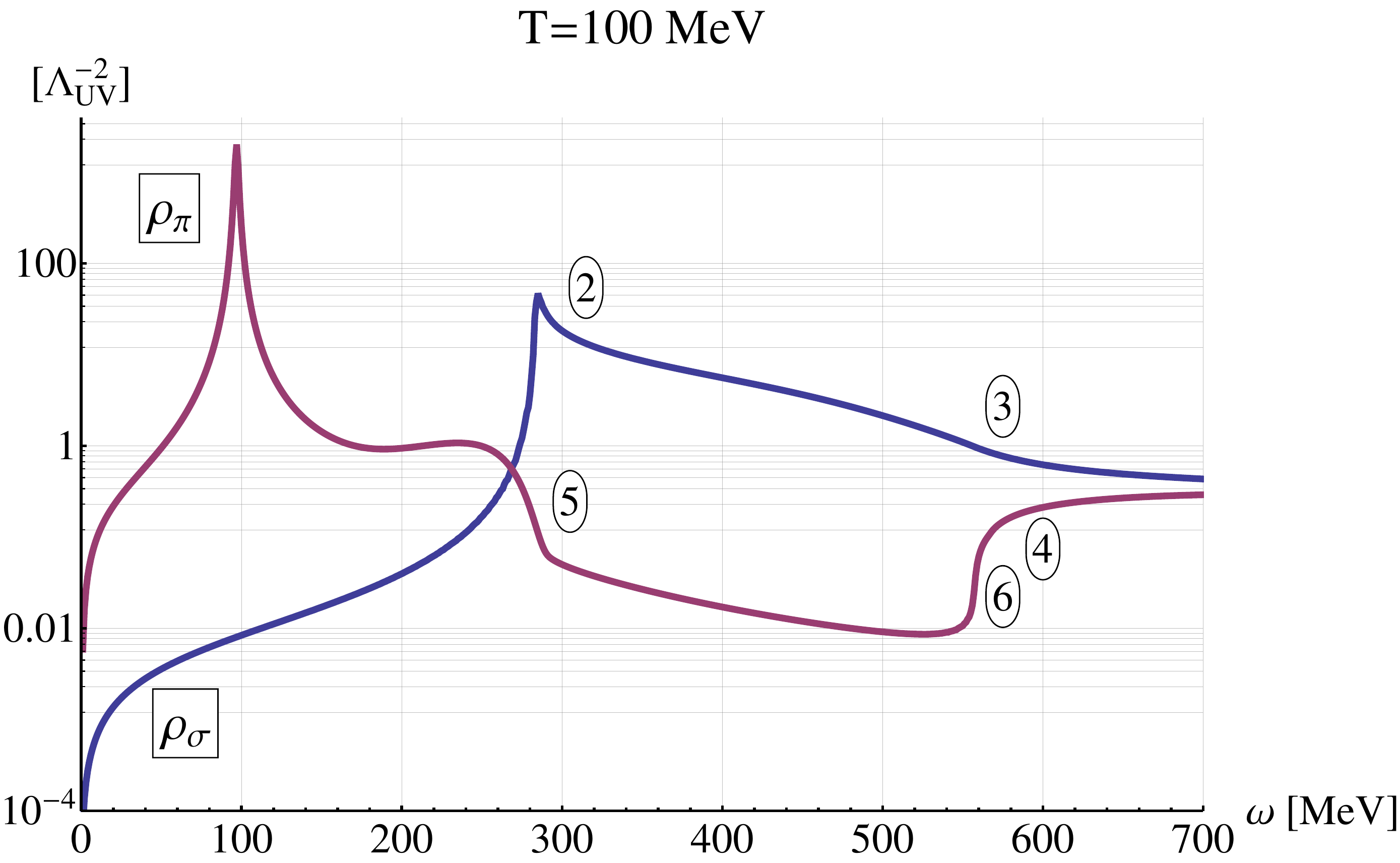}\hspace{5mm}
\includegraphics[width=0.48\columnwidth]{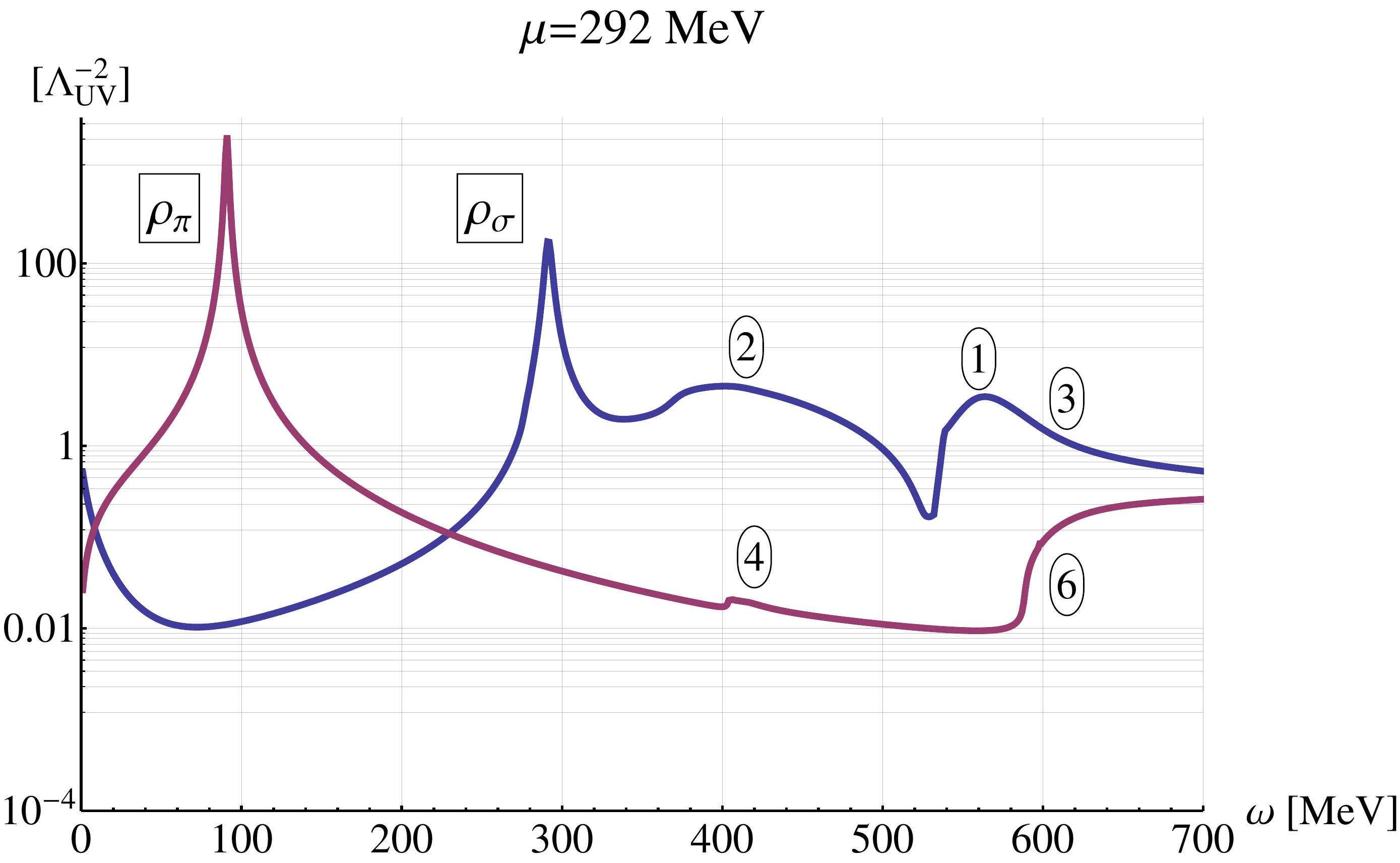}\vspace{3mm}
\includegraphics[width=0.48\columnwidth]{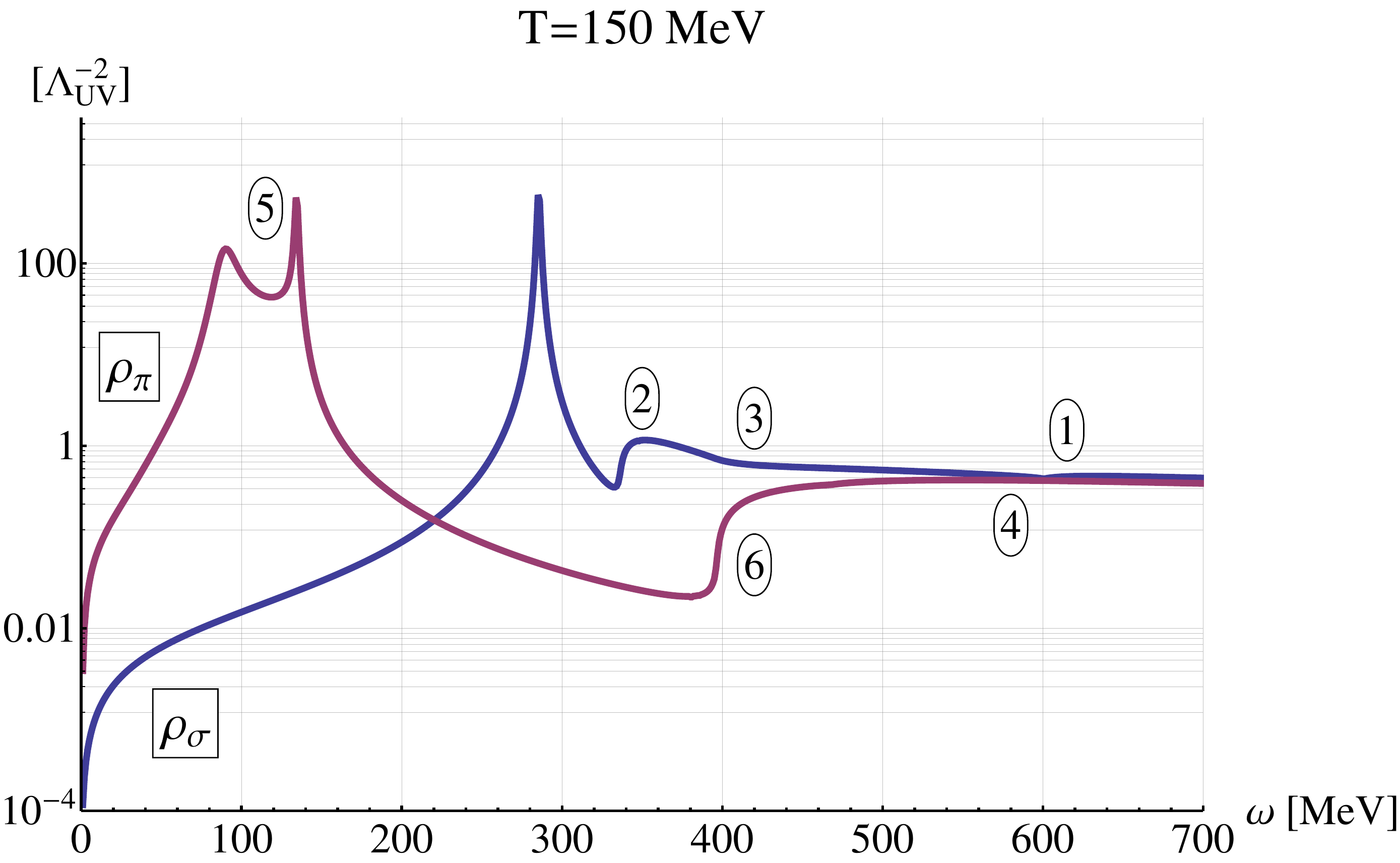}\hspace{5mm}
\includegraphics[width=0.48\columnwidth]{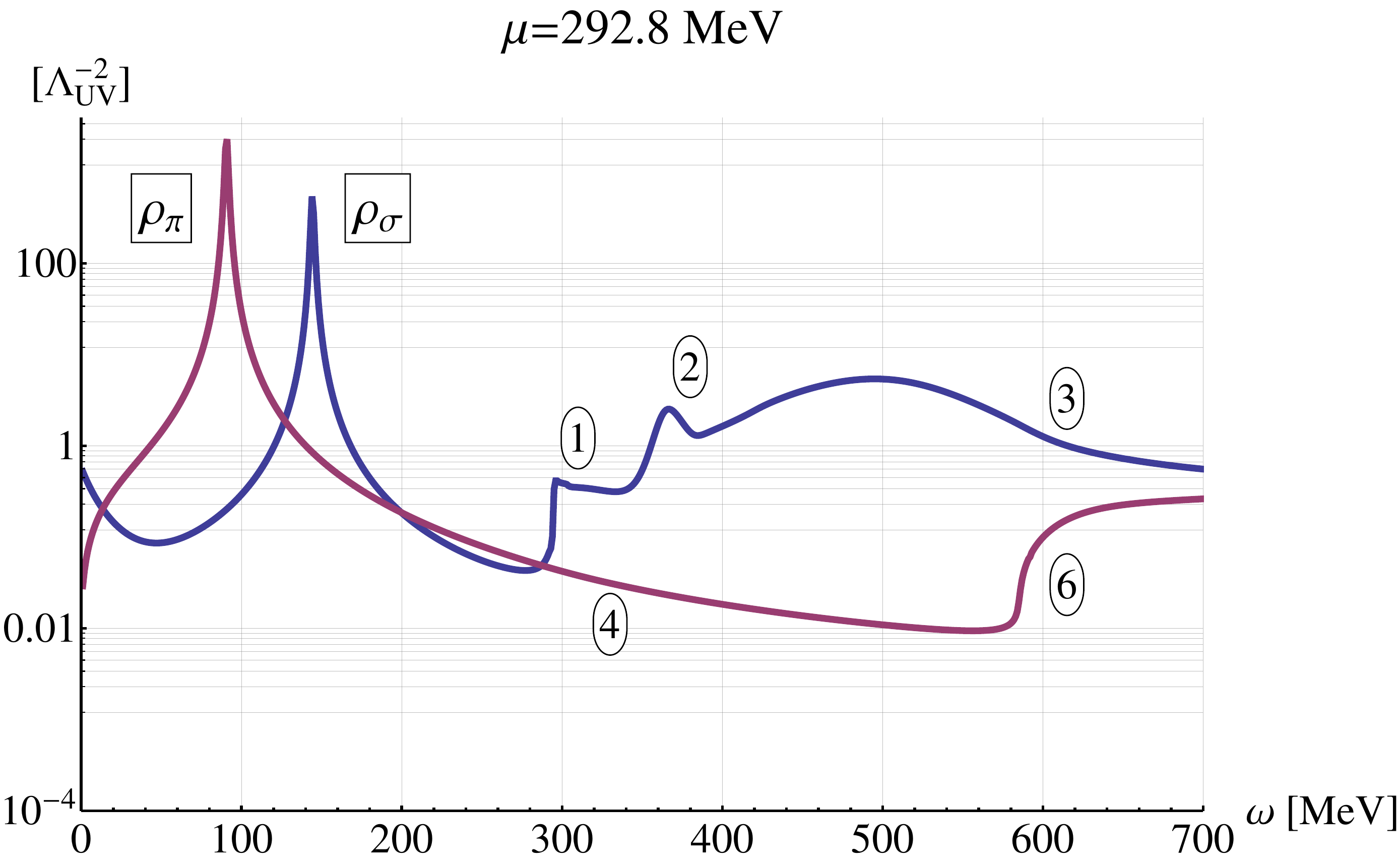}\vspace{3mm}
\includegraphics[width=0.2\columnwidth]{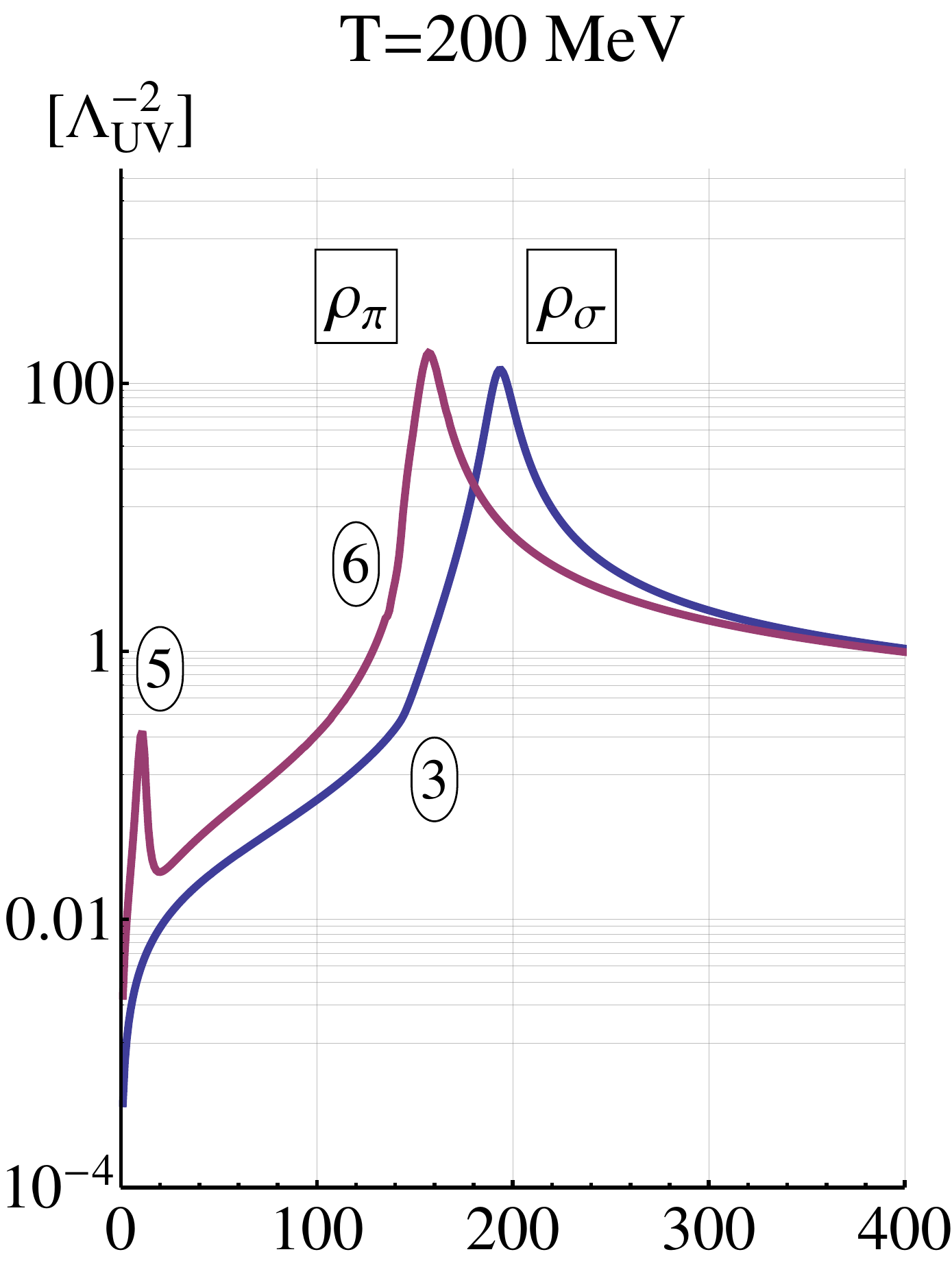}
\includegraphics[width=0.25\columnwidth]{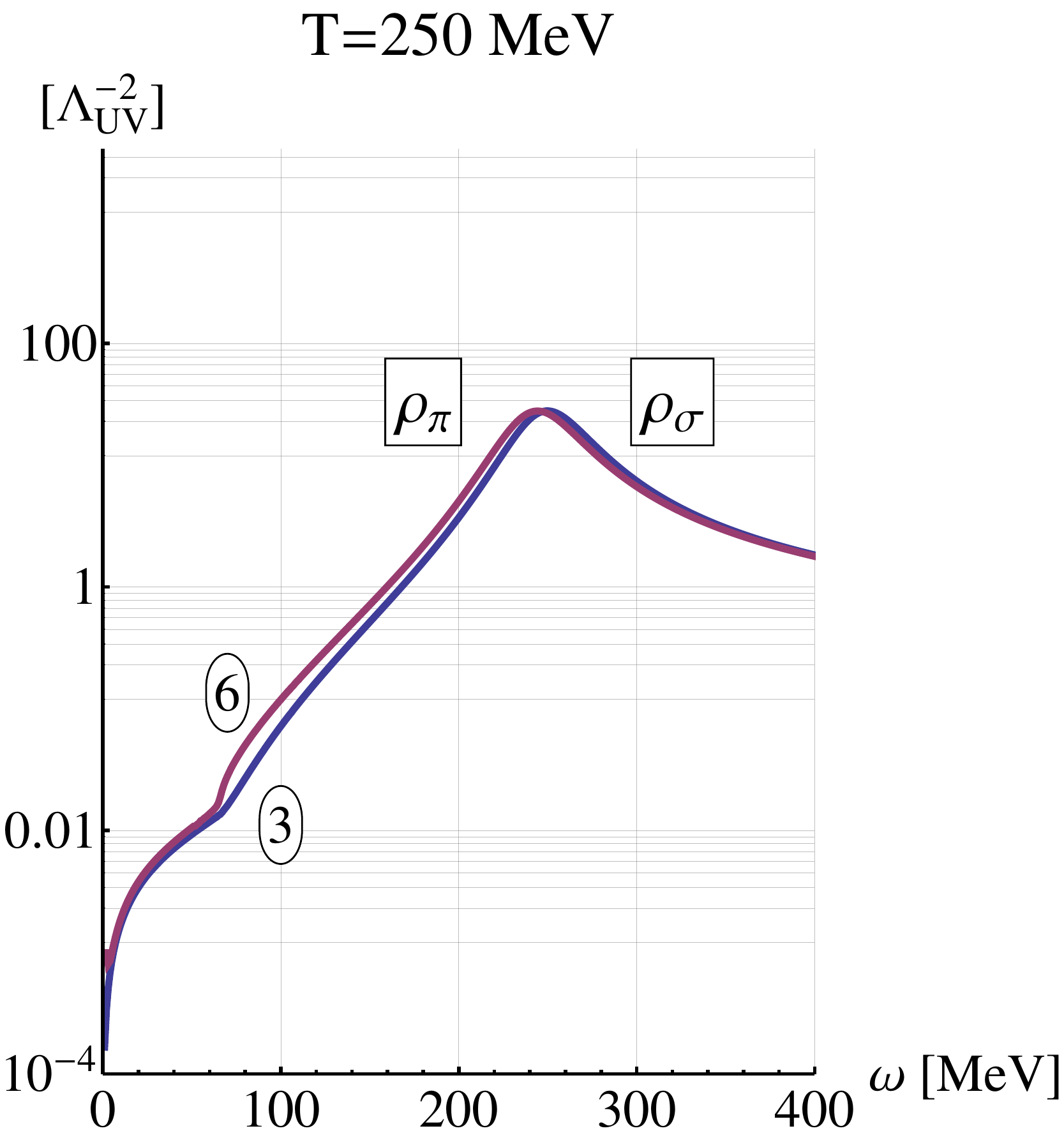}\hspace{8mm}
\includegraphics[width=0.2\columnwidth]{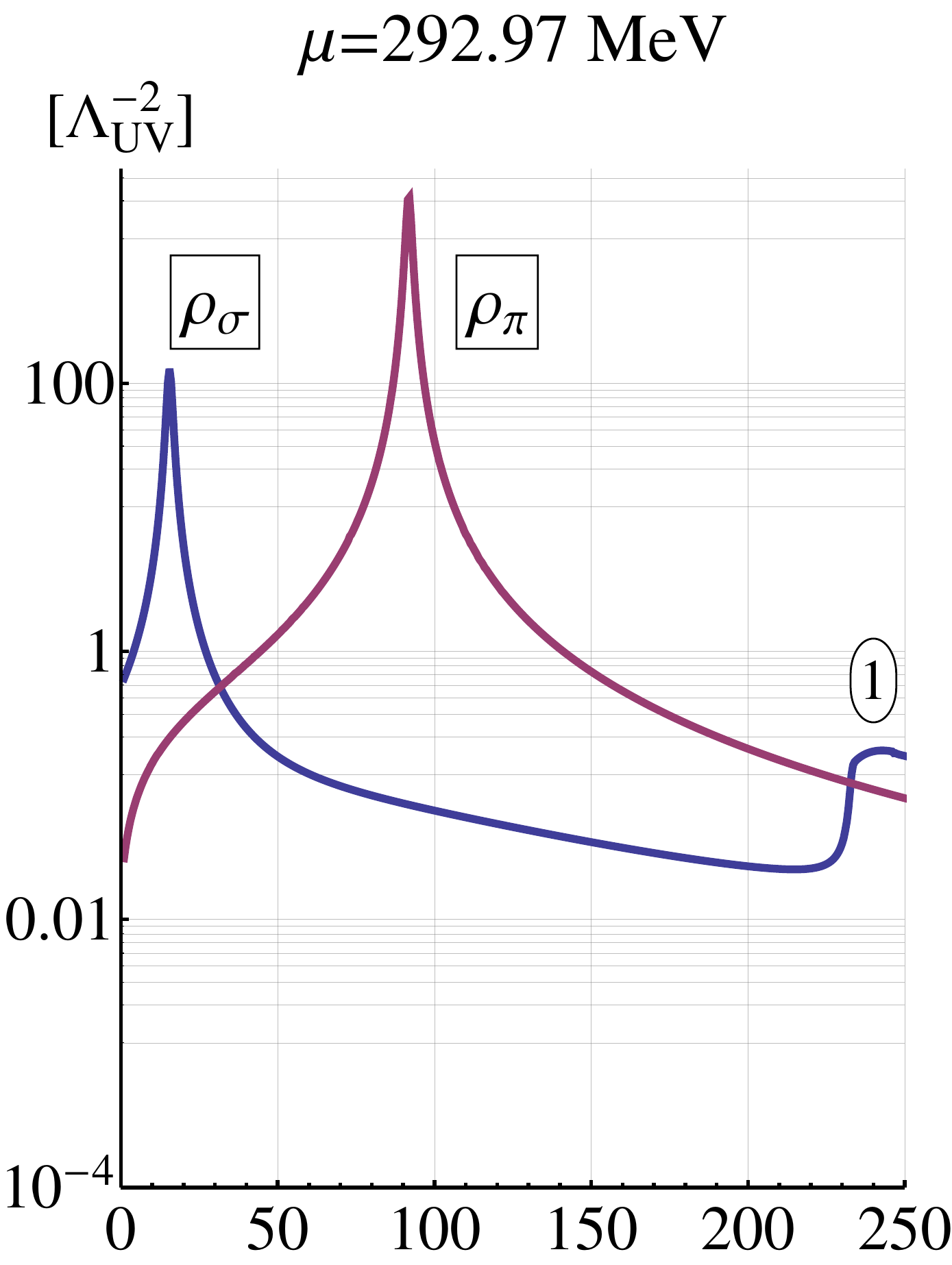}
\includegraphics[width=0.25\columnwidth]{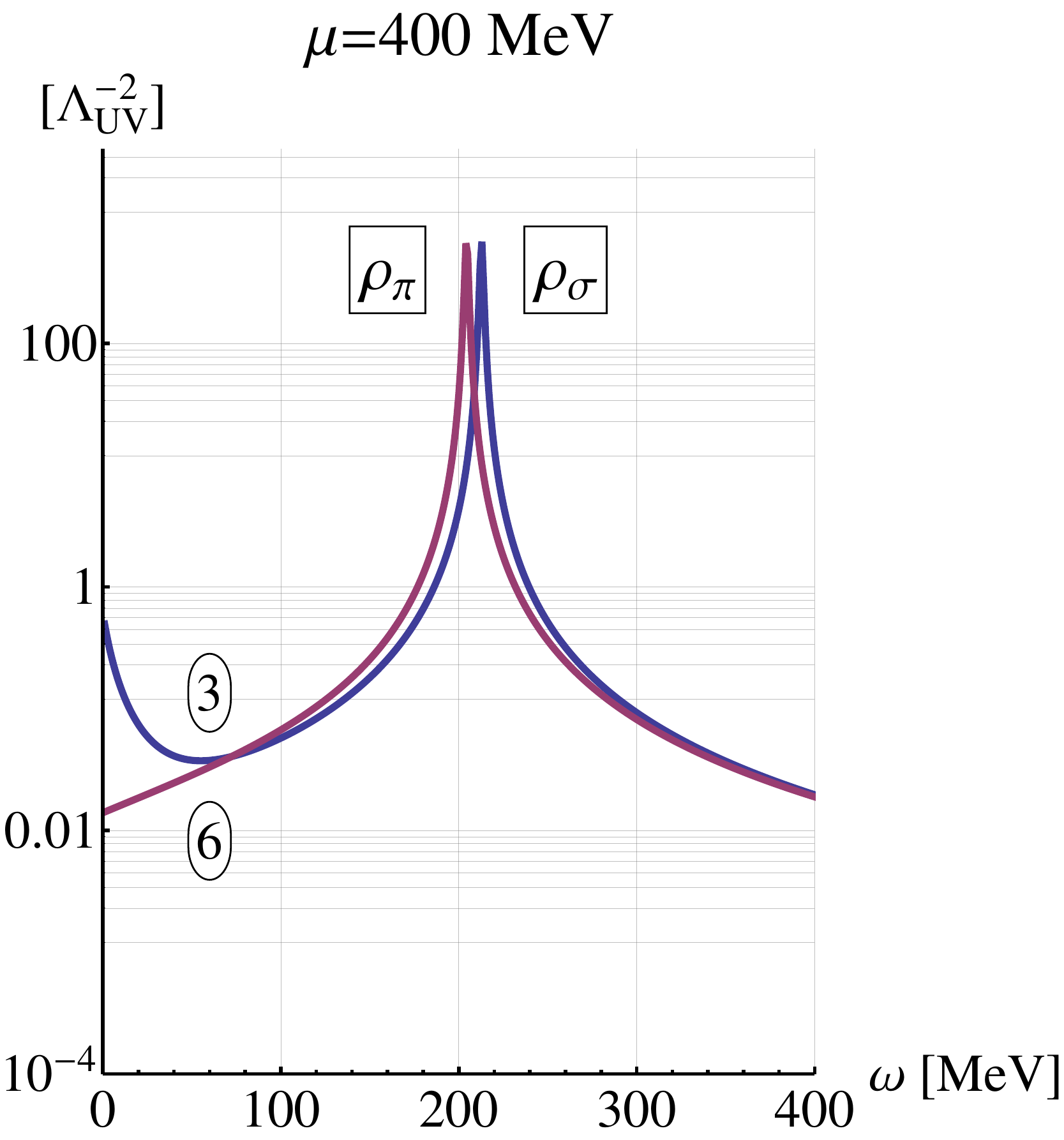}
\caption{(color online) Sigma and pion spectral function from \cite{Tripolt2014}
are shown versus external energy $\omega$ at $\mu=0\,{\rm MeV}$ 
but different $T$ (left column) and at $T=10\,{\rm MeV}$ but different $\mu$ (right column).
Inserted numbers refer to the different processes affecting the spectral functions at corresponding energies. 
1:~$\sigma^* \rightarrow \sigma \sigma$, 
2:~$\sigma^* \rightarrow \pi \pi$, 
3:~$\sigma^* \rightarrow \bar{\psi}\psi$, 
4:~$\pi^* \rightarrow \sigma \pi$, 
5:~$\pi^*\pi \rightarrow \sigma$, 
6:~$\pi^* \rightarrow \bar{\psi}\psi$. 
See text for details.
}
\label{fig:spectralfunctions} 
\end{figure*}

We now turn to the discussion of results for the sigma and pion spectral functions, 
$\rho_\sigma(\omega)$ and $\rho_\pi(\omega)$. They are displayed in Fig.~\ref{fig:spectralfunctions} 
as a function of external energy $\omega$ at different 
values of $T$ and $\mu$ . The inserted numbers refer to different 
processes contributing to the spectral functions via the corresponding diagrams shown in 
Fig.~\ref{fig:flow_Gamma2}. The sigma spectral function is affected by the processes 
${\sigma^*\rightarrow \sigma\sigma}$ (1), ${\sigma^*\rightarrow \pi\pi}$ (2) and 
${\sigma^*\rightarrow \bar{\psi} \psi}$ (3), where primes denote
off-shell correlations with energy $\omega$. The relevant processes
for the pion spectral function are ${\pi^*\rightarrow \sigma\pi}$ (4),
${\pi^*\pi\rightarrow \sigma}$ (5) and  
${\pi^*\rightarrow \bar{\psi} \psi}$ (6). In our truncation with momentum-independent vertices 
the mesonic tadpole diagrams only give rise to $\omega$-independent contributions to the spectral functions.

At $T=10\,{\rm MeV}$ and $\mu= 0\,{\rm MeV}$ the spectral functions closely 
resemble those in the vacuum, observed in the case of the O(4) model, 
with the addition of the quark-antiquark decay channel. When going to higher temperatures, 
the thermal scattering process ${\pi^*\pi\rightarrow \sigma}$,
affects to the pion spectral function for $\omega\leq m_\sigma-m_\pi$. 
Another effect induced by the temperature dependence of 
the meson and quark masses is the emergence of a stable sigma meson at temperatures close to 
the crossover temperature, where neither the decay into two pions nor that into two quarks are 
energetically possible. At $T = 150 \,{\rm MeV}$ we therefore observe a pronounced peak in the sigma 
spectral function at $\omega \approx 280 \,{\rm MeV}$.
When increasing the temperature further, 
the quarks become the lightest degrees of freedom in the system, providing decay channels 
for both the pion and the sigma meson. At very high temperatures
the spectral functions become degenerate,
as to be expected from the progressing restoration of chiral symmetry.

The right column of Fig. \ref{fig:spectralfunctions} shows the sigma and pion 
spectral functions at a fixed temperature of $T=10\,{\rm MeV}$ and different 
values of the quark chemical potential. 
Up to chemical potentials of around 200 MeV the spectral functions remain essentially
unchanged since the ground state is unaffected by the chemical potential below the phase transition
(Silver Blaze property, \cite{Cohen2003}).
When approaching the critical endpoint, however, especially the sigma
spectral function undergoes significant changes.
At $\mu = 292 \,{\rm MeV}$, i.e.\ only about $1 \,{\rm MeV}$ from the
critical endpoint,  
the sigma screening mass has already dropped to about half of its vacuum value, 
leading to a minimal energy for the ${\sigma^*\rightarrow \sigma\sigma}$ decay of 
$\omega\geq 2\,m_\sigma\approx 540\,{\rm MeV}$. In addition, a pronounced sigma 
peak starts to develop at $\omega \approx 290 \,{\rm MeV}$, indicating the 
formation of a stable dynamical sigma meson. 
Even closer to the CEP, at $\mu = 292.8 \,{\rm MeV}$, the threshold for the two-sigma
decay has decreased to $\omega \approx 290 \,{\rm MeV}$ and thus
occurs already at smaller energies  than the ${\sigma^*\rightarrow\pi\pi}$ process. 
At $\mu = 292.97 \,{\rm MeV}$, the sigma meson has become almost massless, as expected
near a second order phase transition. When increasing the chemical potential further, 
the sigma and pion spectral functions become degenerate, similar to the case of high temperatures.

\section{Summary and Outlook}

In this contribution we have presented a tractable scheme for obtaining hadronic spectral
functions at finite temperature and density within the FRG approach. 
The method is based on an analytic continuation from imaginary to real frequencies 
on the level of the flow equations for the pertinent two-point functions. It is thermodynamically 
consistent and symmetry preserving. As a consequence, phase transitions and critical behavior can be 
properly handled. 

We have demonstrated the feasibility of the method by applying it to the O(4) linear-sigma model in the 
vacuum and to the quark-meson model at finite temperature and density. 
Many additional in-medium scattering processes result in a rather complicated 
structure of the spectral functions, clearly exhibiting the critical behavior 
in the vicinity of chiral phase transitions. As chiral symmetry gets restored 
at high temperature and density the spectral function of the pion and the sigma 
meson, as its chiral partner, become degenerate and thus lead to parity doubling.

Apart from extensions to other spectral functions, such as those of the $\rho$ and $a_1$ meson, 
which are of relevance for electromagnetic probes in heavy-ion collisions, there is also the 
exciting possibility to compute transport coefficients in the entire ($T,\mu$)-plane of the phase diagram.\\

\section{Acknowledgements}

We dedicate this article to Gerry Brown, an inspiring physicist and a great human being. 
We would like to thank K. Kamikado for collaboration. This work was supported by the 
Helmholtz International Center for FAIR within the LOEWE initiative of the state of Hesse. 
R.-A.~T. is furthermore supported by the Helmholtz Research School for Quark Matter Studies, 
H-QM, and N.~S. is supported by Grant No. ERC-AdG-290623.


\bibliography{Gerry.bbl} 

\begin{thebibliography}{10}
\expandafter\ifx\csname url\endcsname\relax
  \def\url#1{\texttt{#1}}\fi
\expandafter\ifx\csname urlprefix\endcsname\relax\def\urlprefix{URL }\fi
\expandafter\ifx\csname href\endcsname\relax
  \def\href#1#2{#2} \def\path#1{#1}\fi

\bibitem{Brown1991}
G.~Brown, M.~Rho, {Scaling effective Lagrangians in a dense medium},
  Phys.Rev.Lett. 66 (1991) 2720--2723.
\newblock \href {http://dx.doi.org/10.1103/PhysRevLett.66.2720}
  {\path{doi:10.1103/PhysRevLett.66.2720}}.

\bibitem{Brown2004}
G.~Brown, M.~Rho, {Double decimation and sliding vacua in the nuclear many body
  system}, Phys.Rept. 396 (2004) 1--39.
\newblock \href {http://arxiv.org/abs/nucl-th/0305089}
  {\path{arXiv:nucl-th/0305089}}, \href
  {http://dx.doi.org/10.1016/j.physrep.2004.02.002}
  {\path{doi:10.1016/j.physrep.2004.02.002}}.

\bibitem{Berges:2000ew}
J.~Berges, N.~Tetradis, C.~Wetterich, Nonperturbative renormalization flow in
  quantum field theory and statistical physics, Phys.Rept. 363 (2002) 223--386.
\newblock \href {http://arxiv.org/abs/hep-ph/0005122}
  {\path{arXiv:hep-ph/0005122}}.

\bibitem{Polonyi:2001se}
J.~Polonyi, Lectures on the functional renormalization group method, Central
  Eur.J.Phys. 1 (2003) 1--71.
\newblock \href {http://arxiv.org/abs/hep-th/0110026}
  {\path{arXiv:hep-th/0110026}}, \href {http://dx.doi.org/10.2478/BF02475552}
  {\path{doi:10.2478/BF02475552}}.

\bibitem{Pawlowski:2005xe}
J.~M. Pawlowski, Aspects of the functional renormalisation group, Annals Phys.
  322 (2007) 2831--2915.
\newblock \href {http://arxiv.org/abs/hep-th/0512261}
  {\path{arXiv:hep-th/0512261}}, \href
  {http://dx.doi.org/10.1016/j.aop.2007.01.007}
  {\path{doi:10.1016/j.aop.2007.01.007}}.

\bibitem{Schaefer:2006sr}
B.-J. Schaefer, J.~Wambach, Renormalization group approach towards the qcd
  phase diagram, Phys.Part.Nucl. 39 (2008) 1025--1032.
\newblock \href {http://arxiv.org/abs/hep-ph/0611191}
  {\path{arXiv:hep-ph/0611191}}, \href
  {http://dx.doi.org/10.1134/S1063779608070083}
  {\path{doi:10.1134/S1063779608070083}}.

\bibitem{Kopietz2010}
P.~Kopietz, L.~Bartosch, F.~Schutz, {Introduction to the functional
  renormalization group}, Lect.Notes Phys. 798 (2010) 1--380.
\newblock \href {http://dx.doi.org/10.1007/978-3-642-05094-7}
  {\path{doi:10.1007/978-3-642-05094-7}}.

\bibitem{Braun:2011pp}
J.~Braun, Fermion interactions and universal behavior in strongly interacting
  theories, J.Phys. G39 (2012) 033001.
\newblock \href {http://arxiv.org/abs/1108.4449} {\path{arXiv:1108.4449}},
  \href {http://dx.doi.org/10.1088/0954-3899/39/3/033001}
  {\path{doi:10.1088/0954-3899/39/3/033001}}.

\bibitem{Gies2012}
H.~Gies, Introduction to the functional rg and applications to gauge theories,
  Lect. Notes Phys. 852 (2012) 287--348.
\newblock \href {http://arxiv.org/abs/hep-ph/0611146}
  {\path{arXiv:hep-ph/0611146}}.

\bibitem{Vidberg:1977}
H.~J. {Vidberg}, J.~W. {Serene}, {Solving the Eliashberg equations by means of
  N-point Pad{\'e} approximants}, Journal of Low Temperature Physics 29 (1977)
  179--192.
\newblock \href {http://dx.doi.org/10.1007/BF00655090}
  {\path{doi:10.1007/BF00655090}}.

\bibitem{Jarrell:1996}
M.~Jarrell, J.~Gubernatis, Bayesian inference and the analytic continuation of
  imaginary-time quantum monte carlo data, Physics Reports 269~(3) (1996) 133
  -- 195.
\newblock \href {http://dx.doi.org/10.1016/0370-1573(95)00074-7}
  {\path{doi:10.1016/0370-1573(95)00074-7}}.

\bibitem{Asakawa:2000tr}
M.~Asakawa, T.~Hatsuda, Y.~Nakahara, {Maximum entropy analysis of the spectral
  functions in lattice QCD}, Prog.Part.Nucl.Phys. 46 (2001) 459--508.
\newblock \href {http://arxiv.org/abs/hep-lat/0011040}
  {\path{arXiv:hep-lat/0011040}}, \href
  {http://dx.doi.org/10.1016/S0146-6410(01)00150-8}
  {\path{doi:10.1016/S0146-6410(01)00150-8}}.

\bibitem{Dudal2013}
D.~Dudal, O.~Oliveira, P.~J. Silva, {K\"all\'en-Lehmann spectroscopy for
  (un)physical degrees of freedom. }\href {http://arxiv.org/abs/1310.4069}
  {\path{arXiv:1310.4069}}.

\bibitem{Strodthoff:2011tz}
N.~Strodthoff, B.-J. Schaefer, L.~von Smekal, {Quark-meson-diquark model for
  two-color QCD}, Phys.Rev. D85 (2012) 074007.
\newblock \href {http://arxiv.org/abs/1112.5401} {\path{arXiv:1112.5401}},
  \href {http://dx.doi.org/10.1103/PhysRevD.85.074007}
  {\path{doi:10.1103/PhysRevD.85.074007}}.

\bibitem{Kamikado2013}
K.~Kamikado, N.~Strodthoff, L.~von Smekal, J.~Wambach, {Fluctuations in the
  quark-meson model for QCD with isospin chemical potential}, Phys.Lett. B718
  (2013) 1044--1053.
\newblock \href {http://arxiv.org/abs/1207.0400} {\path{arXiv:1207.0400}},
  \href {http://dx.doi.org/10.1016/j.physletb.2012.11.055}
  {\path{doi:10.1016/j.physletb.2012.11.055}}.

\bibitem{Floerchinger2012}
S.~Floerchinger, {Analytic Continuation of Functional Renormalization Group
  Equations}, JHEP 1205 (2012) 021.
\newblock \href {http://arxiv.org/abs/1112.4374} {\path{arXiv:1112.4374}},
  \href {http://dx.doi.org/10.1007/JHEP05(2012)021}
  {\path{doi:10.1007/JHEP05(2012)021}}.

\bibitem{Tripolt2014}
R.-A. Tripolt, N.~Strodthoff, L.~von Smekal, J.~Wambach, {Spectral Functions
  for the Quark-Meson Model Phase Diagram from the Functional Renormalization
  Group}, Phys.Rev. D89 (2014) 034010.
\newblock \href {http://arxiv.org/abs/1311.0630} {\path{arXiv:1311.0630}},
  \href {http://dx.doi.org/10.1103/PhysRevD.89.034010}
  {\path{doi:10.1103/PhysRevD.89.034010}}.

\bibitem{Wilson1971}
K.~G. Wilson, Renormalization group and critical phenomena. ii. phase-space
  cell analysis of critical behavior, Phys. Rev. B 4 (1971) 3184--3205.
\newblock \href {http://dx.doi.org/10.1103/PhysRevB.4.3184}
  {\path{doi:10.1103/PhysRevB.4.3184}}.

\bibitem{Wilson1974}
K.~Wilson, J.~B. Kogut, {The Renormalization group and the epsilon expansion},
  Phys.Rept. 12 (1974) 75--200.
\newblock \href {http://dx.doi.org/10.1016/0370-1573(74)90023-4}
  {\path{doi:10.1016/0370-1573(74)90023-4}}.

\bibitem{Wetterich:1992yh}
C.~Wetterich, Exact evolution equation for the effective potential, Phys. Lett.
  B301 (1993) 90--94.
\newblock \href {http://dx.doi.org/10.1016/0370-2693(93)90726-X}
  {\path{doi:10.1016/0370-2693(93)90726-X}}.

\bibitem{Kamikado2014}
K.~Kamikado, N.~Strodthoff, L.~von Smekal, J.~Wambach, {Real-Time Correlation
  Functions in the O(N) Model from the Functional Renormalization Group},
  Eur.Phys.J. C74 (2014) 2806.
\newblock \href {http://arxiv.org/abs/1302.6199} {\path{arXiv:1302.6199}},
  \href {http://dx.doi.org/10.1140/epjc/s10052-014-2806-6}
  {\path{doi:10.1140/epjc/s10052-014-2806-6}}.

\bibitem{Baym1961}
G.~Baym, N.~D. Mermin, Determination of thermodynamic green's functions, J.
  Math. Phys. 2 (1961) 232.
\newblock \href {http://dx.doi.org/http://dx.doi.org/10.1063/1.1703704}
  {\path{doi:http://dx.doi.org/10.1063/1.1703704}}.

\bibitem{Landsman1987}
N.~Landsman, C.~van Weert, {Real and Imaginary Time Field Theory at Finite
  Temperature and Density}, Phys.Rept. 145 (1987) 141.
\newblock \href {http://dx.doi.org/10.1016/0370-1573(87)90121-9}
  {\path{doi:10.1016/0370-1573(87)90121-9}}.

\bibitem{Schaefer:2004en}
B.-J. Schaefer, J.~Wambach, The phase diagram of the quark meson model, Nucl.
  Phys. A757 (2005) 479--492.
\newblock \href {http://arxiv.org/abs/nucl-th/0403039}
  {\path{arXiv:nucl-th/0403039}}, \href
  {http://dx.doi.org/10.1016/j.nuclphysa.2005.04.012}
  {\path{doi:10.1016/j.nuclphysa.2005.04.012}}.

\bibitem{Cohen2003}
T.~D. Cohen, {Functional integrals for QCD at nonzero chemical potential and
  zero density}, Phys.Rev.Lett. 91 (2003) 222001.
\newblock \href {http://arxiv.org/abs/hep-ph/0307089}
  {\path{arXiv:hep-ph/0307089}}, \href
  {http://dx.doi.org/10.1103/PhysRevLett.91.222001}
  {\path{doi:10.1103/PhysRevLett.91.222001}}.

\end{thebibliography}

\end{document}